\documentstyle[pra,aps,floats]{revtex}
\newcommand{\be}{\begin{equation}}
\newcommand{\ee}{\end{equation}}  
\newcommand{\ba}{\begin{eqnarray}}
\newcommand{\ea}{\end{eqnarray}}

         \newcommand{\qb}{\begin{equation}}
         \newcommand{\qba}{\begin{eqnarray}}
         \newcommand{\qbas}{\begin{eqnarray*}}
         \newcommand{\qe}{\end{equation}}
         \newcommand{\qea}{\end{eqnarray}}
         \newcommand{\qeas}{\end{eqnarray*}}
         
\input epsf

\begin{document}
\twocolumn[\hsize\textwidth\columnwidth\hsize\csname
@twocolumnfalse\endcsname

\title{Relational modal interpretation for relativistic quantum field theories}
\author{Gyula Bene\cite{email1}}
\address{Institute of Theoretical Physics, E\"otv\"os University,\\
P\'azm\'any P\'eter s\'et\'any 1/A, H-1117 Budapest, Hungary\footnote{z3}}
\date{\today}
\maketitle

\begin{abstract}%
The relational version of the modal interpretation offers both a consistent 
quantum ontology and solution for quantum paradoxes 
within the framework of nonrelativistic quantum mechanics. In the present
paper this approach is generalized for the case of relativistic quantum field
theories. Physical systems are defined as Hilbert spaces. The concept
of the  reduced density matrix is also generalized so that its trace 
may become smaller than one, expressing the possibility of annihilation.
Superselection rules are shown to follow if the whole Universe 
has a definite electric charge, barionic number and leptonic number.  
\end{abstract}

\vskip2pc] 

\narrowtext
\section{Introduction}
Modal interpretations aim at extracting a consistent pysical
picture out of the formalism of quantum mechanics rather than relying on 
assumptions about an {\em a priori} classical world\cite{VanFraassen},
\cite{Dieks}, \cite{Healey} \cite{DieksVermaas}. 
The Dieks-Vermaas version of the modal interpretations \cite{DieksVermaas} utilizes
the Schmidt states, i.e., the eigenstates of the reduced density matrix,
and identifies them with the actually existing, physical states. 
The significance of the Schmidt bases has previously been emphasized 
- in connection with decoherence and Everett's many worlds interpretation 
- by Zeh\cite{Zeh}. The no-go theorem of Vermaas\cite{nogoVermaas} stating the impossibility 
of defining probabilities 
for the simultaneous existence of certain physical states can be understood
within the framework of an interpretation which can be called the relational
version of the modal interpretation\cite{Bene97}.\footnote{This has been 
independently developed and it turned out later that it involves both the 
essential ideas of the modal interpretations and those of the relational 
interpretations.} Relational ideas have a long history. They appeared
first in the original version of Everett's interpretation\cite{Everett},
then, in different forms, in\cite{Mermin} and in \cite{Rovelli}. The essential
idea is that states do not exist in an absolute sense but can only be defined
with respect to another system (or another state\cite{Everett}). This
idea has been implemented in\cite{Bene97} in a way mathematically different
from the previous propositions. The quantum reference systems here contain
the system to be described. The physical states defined in the      
 Dieks-Vermaas interpretation can be identified by the states of a system
with respect to itself (i.e., when the quantum reference system coincides
with the system to be described). The no-go theorem of Vermaas means now
that certain states of different systems that are defined 
with respect to different quantum reference systems cannot be compared, 
not even in principle. This circumstance has been shown to be consistent
with the experimental possibilities which are available according 
to the theory, on the other hand, it clearly goes beyond the usual ontology.
Indeed, one expects that existing things, even if they are defined with 
respect to different reference systems, must somehow be comparable. This
expectation is actually based on classical experience, and its failure
does not violate any well founded physical principle. The quantum ontology
emerging from the relational modal interpretation states that even
the existence of the states cannot be imagined independently of the quantum
reference systems. One cannot think of reality as a big book 
where all the states of any systems with respect
to any quantum reference systems are carefully registered. Such a registration
would readily imply that the simultaneous existence of any states
 can always be checked, i.e., any states are comparable. Precisely this is
impossible. This startling statement of quantum ontology is closely related
to the fact that the state of the whole Universe (this would be the 
``big book'')
does not determine uniquely the state of a subsystem with respect to 
itself, only a set of possible states and their probabilities. On the other 
hand, this fundamental change of the ontology, i.e., of the very concept of 
realism, is necessary in view of Bell's theorem\cite{Bell}. 
By now it is well known that Bell's theorem and the corresponding 
experiments which convincingly support quantum mechanical predictions
\cite{experiments}, \cite{exp2}
imply that at least one fundamental concept should be given up or modified,
either locality, causality or realism\footnote{Sometimes other concepts like 
scientific inference are also questioned.}. Modal interpretations
satisfy all the requirements of locality and causality, 
thus the remaining option
is that the concept of reality should be modified. 
Indeed, it has been shown that 
accepting the above quantum ontology Bell's inequality does not 
follow\cite{Bene00}.
It is instructive to consider the famous Einstein-Podolsky-Rosen (EPR)
reality criterion\cite{EPR} from our point of view. This criterion states that

{\em ``If, without in any way disturbing a system, we can predict 
with certainty (i.e., with probability equal to unity) the value of a physical
quantity, then there exists an element of physical reality
corresponding to this physical quantity.''}

 Now the point is that the value 
of a physical quantity depends on the state of the system, and this state
must be given with respect to some quantum reference system. Thus, the EPR
criterion is valid only if neither the system itself, nor
the quantum reference system is disturbed. But in case 
of the EPR paradox\cite{EPR},
the quantum reference system is disturbed, so the EPR reality criterion is not
applicable and the conclusion about the incompleteness of quantum mechanics
does not follow\cite{Benex}. Note that already Bohr has claimed (albeit using
different arguments) that the concept of realism changes in quantum mechanics
\cite{Bohr}. 

All the above considerations has been done within the framework 
of nonrelativistic quantum mechanics.
In the present paper the relational modal interpretation\cite{Bene97}
is generalized to the case of relativistic field theories. 
Such generalizations of the Dieks-Vermaas version 
of the modal interpretation has already been proposed in Refs.\cite{Dieksrel},
\cite{Cliftonrel}.
In Section II. the concept of the physical systems is given and discussed.
Section III. contains the main result, i.e., the generalized postulates of the 
interpretation. These replace von Neumann's measurement postulates,
thus making quantum theory self consistent, i.e., removing the necessity
of an {\em a priori} classical background.  
In Section IV. the possible origin of the superselection rules is discussed.
In the concluding Section V. a summary of the results is given.   
\section{The concept of the physical systems}

In nonrelativistic quantum mechanics physical systems might be specified
by the particles they contain. This definition becomes unsatisfactory, 
however, when macroscopic systems are concerned. Indeed, 
a description of a macroscopic system should contain its structure 
as well, which is not included if only its constituent particles are given.
Moreover, this structure is much more important than the precise
number of the particles. A straightforward possibility is to specify
a system by the collection of states which correspond to the structure and
functionality of that system. These states may even contain a different number
of particles. Certainly, the superposition principle must be respected 
(at least to an extent allowed by the superselection rules), thus 
arbitrary linear combinations of these states are also allowed. This makes
the collection of the states a vector space. As the scalar product of these
states are defined as usually, we have an Eucledian vector space. Finally,
the completion of this space gives rise to a Hilbert space, which is much
narrower than the total Hilbert space of all the constituent particles.
E.g., when constructing the Hilbert space of a measuring device as described 
above, one does not include states which correspond to a destructed device.
This construction can be equally well applied in case 
of relativistic field theories. In that case states are given in Fock space,
thus typically contain superpositions of states with different occupation 
numbers.
In the nonrelativistic case interacting systems usually 
can be chosen such that they preserve their identity
during the interaction, while this is in general impossible in the
relativistic case. Mathematically, this means that in the
nonrelativistic case the interaction 
moves the state of the composite system
 within the direct product of the subsystems' Hilbert spaces, while
in the relativistic case the state may leave the direct product space
during the interaction. Note that this situation can appear 
in the nonrelativistic case as well, e.g., if a measuring device is destroyed
by a too hard interaction (say, a too low measuring range has been set), 
the final state of the composite system can be outside of the direct product
space. In the relativistic case this situation is typical which means that 
a system can disappear. This means that the direct product of the
subsystems' Hilbert spaces is just a subspace of the composite system's
Hilbert space. Certainly, for the description
of the interactions one has to choose such a Hilbert space (i.e., such
a composite system) which is broad enough to accomodate the state during
the whole time evolution. Such a system can be called isolated (as it does
not interact with the rest of the world). Strictly speaking, 
there is only one such system: the whole Universe itself.

Sometimes we may assume
that in the absence of interactions with other systems 
time evolution moves the state of the system within its original Hilbert space.
Even this condition can be released, as it is reasonable 
in case of open systems
like living beings. Indeed, a living being would die at once in the absence of
interactions and thus would leave the Hilbert space which defines
it on the basis of its normal functionality.  

\section{Postulates}
  
Once physical systems are defined mathematically as suitable Hilbert spaces,
the next technical problem is how to give the state of a system with respect
to another (broader) one. Here we follow Ref.\cite{Bene97} and
make the necessary generalizations to get consistent rules.

Let us denote the physical system to be described by $A$ and the
reference system by $R$. The state of $A$ with respect to $R$
will be denoted by $\hat \rho_A(R)$.
\vskip0.5cm
  {\em Postulate 1. The system A to be described is a subsystem of the
reference system R.}
\vskip0.5cm
As the systems are now defined as Hilbert spaces, the concept of the subsystem
needs to be suitably defined as well. 
\vskip0.5cm
{\em Definition 0.  $A$ is a
subsystem of $R$ if there exists another Hilbert space $B$ so that}
\be
A\otimes B \subseteq R
\ee
{\em The broadest system $B$ is called the complementer system of $A$
  (with respect to $R$) and is denoted by $R/A$.}
\footnote{Because the construction of $B$ is indeed some kind of division.}
\vskip0.5cm
Note that the reference system may coincide with the system to
be described ($A=R$). In such a case we speak about an {\em
internal state}.
\vskip0.5cm
{\em Definition 1. $\hat \rho_A(A)$ is called the internal state
of $A$.}
\vskip0.5cm
{\em Postulate 2. The state $\hat \rho_A(R)$ is a positive definite,
Hermitian
operator with a trace not larger than one, acting on $A$.}
\vskip0.5cm
Note that the fact that the trace can be smaller than one 
is related to the possibility of annihilation.
\vskip0.5cm
{\em Postulate 3. The internal states $\hat \rho_A(A)$ are always
projectors, i.e., $\hat \rho_A(A)=|\psi_A><\psi_A|$.}
\vskip0.5cm
In what follows these projectors will be identified with the
corresponding wave functions $|\psi_A>$ (as they are uniquely
related, apart from a phase factor). In accordance with {\em Postulate 2}
we assume that the internal states are normalized to unity.
\vskip0.5cm
{\em Postulate 4. The state of a system $A$ with respect to the
reference system $R$ (denoted by $\hat \rho_A(R)$) is the reduced
density matrix of $A$ calculated from the internal state of $R$,
i.e.
\begin{eqnarray}
\hat \rho_A(R)=Tr_{R/A} \left(\hat \rho_R(R)
\right)\quad,
 \label{e8}
\end{eqnarray}
where $R/ A$ stands for the subsystem of $R$
complementer to $A$ (cf. Definition 0).}
\vskip0.5cm
Eq.(\ref{e8}) can be expressed in other forms, too. If $|\xi_{A,j}>$
is a complete, orthonormed basis in $A$ and
$|\xi_{R/A,j}>$
is a complete, orthonormed basis in $R/A$, then
\begin{eqnarray}
\hat \rho_A(R)\mbox{\hspace{7cm}}\nonumber\\
=\sum_{j,k}|\xi_{A,j}>\left(Tr_{R} \left(|\xi_{A,k}>
<\xi_{A,j}|\hat \rho_R(R)
\right)\right)<\xi_{A,k}|\nonumber\\
=\sum_{j,k}|\xi_{A,j}>\left(Tr_{R/ A}<\xi_{A,j}|\hat \rho_R(R)|\xi_{A,k}>
\right)<\xi_{A,k}|\label{e8a1}
\end{eqnarray}
or
\begin{eqnarray}
\hat \rho_A(R)=\sum_{j,k,l}|\xi_{A,j}><\xi_{A,k}|\mbox{\hspace{2.5cm}}\nonumber\\
\times\left(<\xi_{A,j}|\otimes <\xi_{R/A,l}|\right)|\psi_R>\nonumber\\
\times<\psi_R|\left(|\xi_{R/A,l}>\otimes |\xi_{A,k}>\right)
 \label{e8a}
\end{eqnarray}
One can see that {\em Postulate 4} is consistent with
{\em Postulate 2} and {\em Postulate 3}. Because $A\otimes (R/A)$
is not necessarily equal to $R$, Eq.(\ref{e8a}) readily implies that 
$Tr_A \hat \rho_A(R)$ can be
smaller than unity:
\begin{eqnarray}
Tr_A \hat \rho_A(R)=\sum_{j}Tr_{R/ A}<\xi_j|\hat \rho_R(R)|\xi_j>
\le 1\quad,
 \label{e8b}
\end{eqnarray}

\vskip0.5cm
As in the nonrelativistic case, we introduce the notion of {\em isolated } and
{\em closed systems}.
\vskip0.5cm
{\em Definition 2. An isolated system is
such a system that has not been interacting
with the outside world. A closed system
is such a system that is not interacting with any other
system at the given instant of time
(but might have interacted in the past).}
\vskip0.5cm
An isolated system can be
described by a wave function, and, moreover, this wave function
always occurs as a factor in the internal state of any
broader system. Thus we set
\vskip0.5cm
{\em Postulate 5. If $I$ is an isolated system then its state is
independent of the reference system $R$:}
\begin{eqnarray}
\hat \rho_I(R)=\hat \rho_I(I)\quad.
 \label{e9}
\end{eqnarray}
\vskip0.5cm
Note that if only the whole universe 
can be considered an isolated system, then {\em Postulate 5} becomes
superfluous.

\vskip0.5cm
{\em Postulate 6. If the reference system $R=I$ is an isolated one
then the state $\hat \rho_A(I)$ commutes with the
internal state $\hat \rho_A(A)$.}
\vskip0.5cm
This means that the internal state of $A$ coincides with
one of the eigenstates of $\hat \rho_A(I)$.
Note that usually there is no one-to-one correspondence
between states with respect to different reference systems, thus
one cannot tell (knowing $\hat \rho_A(I)$) which eigenstate
corresponds to $\hat \rho_A(A)$). In what follows,
these eigenstates $|\phi_{A,j}>$ will play an important role.
They will be identified with the corresponding projector
$\hat \pi_{A,j}=|\phi_{A,j}><\phi_{A,j}|$, and we shall call
them the {\em possible internal states}, as they constitute
the set of those internal states of $A$ that are compatible
with $\hat \rho_A(I)$. (Of course, at a given time only one
of these states exists {\em with respect to $A$}.) For further
reference,
we set
\vskip0.5cm
{\em Definition 3. The possible internal states are the eigenstates
of
$\hat \rho_A(I)$ provided that the reference system $I$ is an
 isolated one.}
\vskip0.5cm

 An important property of the possible internal states is their 
connection with the Schmidt decomposition:
\vskip0.5cm
{\em Proposition 1. If $A$ and $B$ are two disjointed physical
 systems (i.e.,
they have no common subsystems) with possible internal states
$|\phi_{A,j}>$ and $|\phi_{B,j}>$, respectively, and the joint
system $A\otimes B$ is an isolated one, then the internal state of
$A\otimes B$ can be written as}
\begin{eqnarray}
|\psi_{A\otimes B}>=\sum_j c_j |\phi_{A,j}>\otimes |\phi_{B,j}>\quad.
 \label{e10}
\end{eqnarray}
\vskip0.5cm

Note that in the relativistic case 
the composite system suitable for the description
of the interaction of $A$ and $B$ can be broader than $A\otimes B$.
Then, certainly, the states of that composite system $C$ contain an
additional term, namely,
\begin{eqnarray}
|\psi_{C}>=\sum_j c_j |\phi_{A,j}>\otimes |\phi_{B,j}>+|\tilde \psi>\quad.
 \label{e10a}
\end{eqnarray}
where $|\tilde \psi>$ is orthogonal to all the states 
$|\phi_{A,j}>\otimes |\phi_{B,k}>$.

\vskip0.5cm
{\em Postulate 7. If $I$ is an isolated system, then the
probability $P(A,j)$ that the
eigenstate $|\phi_{A,j}>$
of $\hat \rho_A(I)$ coincides with $\hat \rho_A(A)$
is given by the corresponding eigenvalue $\lambda_j$.}

Due to the
normalization of the internal states the sum of these eigenvalues 
is not larger than unity.
\vskip0.5cm
{\em Postulate 8. The result of a measurement is contained
unambigously in the internal state of the measuring device.}
\vskip0.5cm
{\em Postulate 9. If $A$ and $B$ are two disjointed physical systems
with possible internal states
$|\phi_{A,j}>$ and $|\phi_{B,j}>$, respectively, and both systems
are contained in the isolated reference system $I$, then the joint
probability that $|\phi_{A,j}>$
coincides with the internal state of $A$ and at the same time
$|\phi_{B,k}>$
coincides with the internal state of $B$ ($j,k=1,2,3,...$) is given
by
\begin{eqnarray}
P(A,j,B,k)=Tr_{A\otimes B} \left(\hat \pi_{A,j} \hat \pi_{B,k} \hat
\rho_{A\otimes B}(I)
\right)\quad,  \label{e16}
\end{eqnarray}
where

$\hat \pi_{A,j}=|\phi_{A,j}><\phi_{A,j}|,\quad
\hat \pi_{B,k}=|\phi_{B,k}><\phi_{B,k}|\quad$.}
\vskip0.5cm

Due to the completeness of the possible internal states
$\sum_k \hat \pi_{B,k}=1$ holds, therefore $\sum_k
P(A,j,B,k)=P(A,j)$, like in the nonrelativistic case.

More generally, we can consider the joint probabilities for
more than two disjointed physical systems. Then we set
\vskip0.5cm
{\em Postulate 10. If there are $n$ disjointed physical systems,
denoted by
\hfill\break
$A_1, A_2, ... A_n$, all contained in the isolated reference
system $I$ and
having the
possible internal states
$|\phi_{A_1,j}>, |\phi_{A_2,j}>,...,|\phi_{A_n,j}>$, respectively,
then the joint
probability that $|\phi_{A_i,j_i}>$
coincides with the internal state of $A_i$ ($i=1,..n$)
is given by
\begin{eqnarray}
P(A_1,j_1,A_2,j_2,...,A_n,j_n)\mbox{\hspace{3cm}}\nonumber\\
=Tr_{A_1\otimes A_2\otimes ...\otimes A_n} [\hat \pi_{A_1,j_1}
\hat \pi_{A_2,j_2}
...\hat \pi_{A_n,j_n}\nonumber\\
\times
\hat \rho_{A_1\otimes A_2\otimes ...\otimes A_n}(I)]\;,\label{e17}
\end{eqnarray}
where $\hat \pi_{A_i,j_i}=|\phi_{A_i,j_i}><\phi_{A_i,j_i}|$.}
\vskip0.5cm
Note that $n=1$ and $n=2$ correspond to {\em Postulate 7} and
{\em Postulate 9},
respectively.
\vskip0.5cm
Finally, we require that
{\em Postulate 11. The internal state $|\psi_C>$ of a closed system
$C$
satisfies the time dependent Schr\"odinger equation}
\begin{eqnarray}
i\hbar \partial_t |\psi_C>=\hat H |\psi_C> \quad.
\label{e28}
\end{eqnarray}
\vskip0.5cm
Here $\hat H$ stands for the Hamiltonian. Here we have used the Schr\"odinger
picture.

\section{Superselection rules}

As a check of the consistency of the present approach I show here
that superselection rules follow for any system if they are valid for the 
whole universe. According to {\em Postulate 6.}, it is enough to show that
the state of a system with respect to the whole universe is such a density
matrix that is diagonal in the electric charge, barionic and leptonic
number. Let us apply {\em Postulate 4.} and Eq.(\ref{e8a}) 
to calculate this state. Here we can
choose - without restricting the generality - the states $|\xi_{A,j}>$, 
$|\xi_{R/A,k}>$
to be charge eigenstates. Now it is clear that if the state $|\psi_R>$ 
is a charge eigenstate, all those terms in Eq.(\ref{e8a}) vanish where
the states $|\xi_{A,j}>$ and $|\xi_{A,k}>$ correspond to different charges.
This is because charge is an additive conserved quantity.
Thus, the state (\ref{e8a}) is indeed diagonal in the charge. The statement
can be proven in the same way for the case of barionic and leptonic number.
     
\section{Summary and conclusion}

As we have seen the postulates of the relational modal interpretation
can be generalized for the case of relativistic field theories.
As a mathematical description of physical systems 
Hilbert spaces has been constructed starting 
from state vectors which express the structure and functionality
of the system. The postulates of Ref.\cite{Bene97} have been 
generalized accordingly. Note that the present formalism 
offers a useful generalization of the previous approach 
even within the framework of the nonrelativistic case. 
In the relativistic case 
 the trace of the states is usually 
smaller than unity when the quantum
reference system is broader than the system to be described.   
The postulates are consistent and, as a result of using 
the trace and eigenvalue equations as basic operations, they are also
independent of the representation.
Finally, it has been demonstrated that the postulates accomodate
the superselection rules in a consistent way.  

\section{Acknowledgements}
Several enlightening discussions with  
Andreas Bringer, Dennis Dieks, Gert Eilenberger, Michael Eisele, G\'eza Gy\"orgyi,
Frigyes K\'arolyh\'azy, Hans Lustfeld, Roland Omnes, Zolt\'an Perj\'es, 
L\'aszl\'o Szab\'o, Mikl\'os R\'edei and
 Pieter Vermaas are gratefully acknowledged.

This work has been partially supported by the Hungarian Aca\-demy of
 Sciences
 under Grant No. OTKA T 029752, T 031 724  and the J\'anos Bolyai Research Fello
wship.

\end{document}